\documentclass[apl,reprint,floatfix]{revtex4-1}
\usepackage[text={6.5in,9.0in},centering]{geometry}
\usepackage{graphicx}
\usepackage{amsmath,amssymb}
\usepackage{subfigure}
\usepackage{bm}
\usepackage{color}
\usepackage[normalem]{ulem}

\newcommand{\red}[1]{\textcolor{red}{{#1}}}

\newlength{\figwidth}
\newcommand{\fref}[1]{Fig.\,\ref{#1}}
\newcommand{\tref}[1]{Table\,\ref{#1}}
\newcommand{\eref}[1]{Eq.\,(\ref{#1})}

\newcommand{\cref}[1]{Ref.\,\cite{#1}}

\newcommand{\ie}{{\it i.e.}\! }
\newcommand{\eg}{{\it e.g.}\! }

\newcommand{\etal}{{\it et al.}\! }

\renewcommand{\ss}{\text{ss}}
\newcommand{\ff}{\text{ff}}
\renewcommand{\sf}{\text{sf}}
\newcommand{\jump}[1]{{[\! [{#1}]\! ]}}
\newcommand{\E}{{\Psi}}

\newcommand{\Fc}{\mathcal{F}}

\newcommand{\Fb}{\mathbf{F}}
\newcommand{\Eb}{\mathbf{E}}

\newcommand{\Nb}{\mathbf{N}}
\newcommand{\Pb}{\mathbf{P}}

\newcommand{\Ib}{\mathbf{I}}

\newcommand{\Jb}{\mathbf{J}}

\newcommand{\xb}{\mathbf{x}}

\newcommand{\ub}{\mathbf{u}}
\newcommand{\Xb}{\mathbf{X}}

\newcommand{\del}{\boldsymbol{\partial}}

\newcommand{\eshelby}{\mathbf{S}}
\newcommand{\chib}{\boldsymbol{\chi}}

\newcommand{\dA}{\mathrm{d}A}

\begin{document}
\title{Atomic-scale interaction of a crack and an infiltrating fluid}
\author{ R.E. Jones$^*$, William C. Tucker, J.M. Rimsza, and L.J. Crisenti } 
\affiliation{Sandia National Laboratories, P.O. Box 969, Livermore, CA 94551, USA}
\begin{abstract}
In this work we investigate the Orowan hypothesis, that decreases in surface energy due to surface adsorbates lead directly to lowered fracture toughness, at an atomic/molecular level.
We employ a Lennard-Jones system with a slit crack and an infiltrating fluid, nominally with gold-water properties, and explore steric effects by varying the soft radius of fluid particles and the influence of surface energy/hydrophobicity via the solid-fluid binding energy.
Using previously developed methods, we employ the $\Jb$-integral to quantify the sensitivity of fracture toughness to the influence of the fluid on the crack tip, and exploit dimensionless scaling to discover universal trends in behavior.
\end{abstract}
\maketitle

\section{Introduction} \label{sec:introduction}

Environmental influences, such as metal oxidation, salt-water corrosion, and fracking \cite{lawn1981mechanics,thouless1990stress,sawada1986influence,chen1980fracture,wong1987microindentation,van1988fluorine}, on fracture are ubiquitous in engineered and natural environments.
Even without significant and complex chemistry, it is widely understood that the cleavage behavior of cracks is controlled by atomic forces and that the characteristic dimensions of crack tips are on the scale of fluid molecules.
A basic tenet of fracture mechanics and propagation is that the fracture toughness of brittle materials is directly related to the energetic cost of forming new surfaces.  
In an early development of fracture mechanics, Orowan \cite{orowan1944fatigue} postulated that the decrease in surface energy due to adsorption of environmental species would decrease effective fracture toughness.
The manifestation of surface energy can involve complex phenomenon at the atomic level, \eg surface reconstructions \cite{baski1997structure} and hydroxylation \cite{rimsza2017surface}. 
Also, the access of fluid to confined spaces like the crack tip is governed by capillary forces, fluid surface tension, and solid-fluid adhesion, all of which have a collective molecular origin.

The interplay of an infiltrating fluid with a crack can be complex.
In fact, Michalske and Freiman \cite{michalske1982molecular} proposed a multistep process of bond breaking in cracked silica exposed to water where, initially, water assists with silica bond breaking, and the end-state energy is affected by surface hydroxylation.
Furthermore, Fisk and Michalske \cite{fisk1985laser} experimentally demonstrated that surface adsorption is an important aspect of chemically assisted fracture.
Michalske \etal \cite{michalske1987steric} also presented experimental evidence of steric exclusion of reactive species by the atomic dimensions of the crack tip.
See Chap. 5 of Lawn \etal\cite{Lawn1993} for a detailed discussion of chemical effects on fracture.

In this study, we employ a Lennard-Jones (LJ) model of a crack infiltrated by a fluid together with dimensionless scaling to remove some of the complexities, \eg surface reconstruction, multi-step chemical reactions and multiple pathways,  that obscure what we assume to be the main contributors to fracture toughness: surface energy, steric considerations and fluid pressure.

\section{Theory} \label{sec:theory}

Here, we briefly review both the theory of configurational forces that connects the $\Jb$-integral to fracture toughness, and the linear elastic solutions for a perfect slit crack with a fluid at pressure.

The $\Jb$-integral is the generalized force energy-conjugate to the generalized displacement defined by the crack tip propagation.
This energy conjugacy is based on the potential energy of an elastic body with a pre-existing crack and accounts for configuration changes due to crack propagation which changes reference configuration of the elastic body.
(The full derivation is subtle, refer to \cref{jones2018atomic} and the classical references therein.)
The key is that the potential energy $\E$ exhibits an explicit dependence on the location of the crack and hence the reference positions $\Xb$, and the derivative of the potential energy with respect to the explicit dependence on $\Xb$ is identified with the divergence of the Eshelby stress $ \eshelby \equiv \E \Ib - \Fb^T \Pb $ which is composed of the free energy density $\E$, the deformation gradient $\Fb \equiv \del_\Xb \chib_t$, and the (first Piola-Kirchhoff) stress $\Pb = \del_\Fb \E$ fields. 
Here, $\xb = \chib_t(\Xb)$ is the motion of the body.
The $\Jb$-integral is defined as the divergence of the Eshelby stress, 
\begin{equation} \label{eq:J-integral} 
\Jb \equiv \int_C \eshelby \Nb \, \dA 
    = \int_C \left( \E \Ib - \Fb^T \Pb \right) \Nb \dA,
\end{equation}
in a region enclosed by $C$ with outward normal $\Nb$.

Linear elastic fracture mechanics (LEFM) provides an analytical solution for tensile (mode I) loading of a semi-infinite slit crack. 
The in-plane displacement field is:
\begin{align} \label{eq:uLEFM}
\ub(R,\theta)
= \frac{K}{2 \mu} \sqrt{\frac{R}{2 \pi}} 
\Bigl[&
\left(\kappa\!-\!1\!+\!2\sin^2\frac{\theta}{2}\right)
 \cos\frac{\theta}{2} \Eb_1 \\
+&
\left(\kappa\!+\!1\!-\!2\cos^2\frac{\theta}{2}\right)
 \sin\frac{\theta}{2} \Eb_2 
\Bigr]  \nonumber
\end{align}
in terms of the shear modulus $\mu$, $\kappa=3-4\nu$ (with $\nu$ being Poisson's ratio), the in-plane polar coordinates $(R,\theta)$, and the Cartesian coordinate vectors $\Eb_1$ and $\Eb_2$.
With this plane strain solution, \eref{eq:J-integral} reduces to $ \Jb \equiv J \Eb_1 = \frac{K^2}{E^*} \Eb_1 $ where $ E^* = \frac{2\mu}{1-\nu}$.
The critical value $ J_c = 2 \gamma $, where $\gamma$ is the surface energy, is associated with the energetic cost of creating new surfaces along the crack.
Particularly relevant to the influence of an infiltrating fluid on fracture, pressure on the crack faces augments $K$-loading (\eref{eq:uLEFM}) by $\Delta K = \frac{2}{\pi} p \sqrt{2\pi w}$, where $w$ is the extent of pressure $p$ loading starting from the crack tip \cite[Sec. 3.7]{tada1973stress}.
External pressure does not change the intrinsic toughness $K_c = \sqrt{J_c E^*}$ of the material; however, with pressure, the apparent $J$ becomes $J = \frac{(K+\Delta K)^2}{E^*}$ where $K$ remains the scaling in the mechanical loading \eref{eq:uLEFM}.

\section{Method} \label{sec:method}

\begin{figure}[h!]
\centering
{\includegraphics[width=\figwidth]{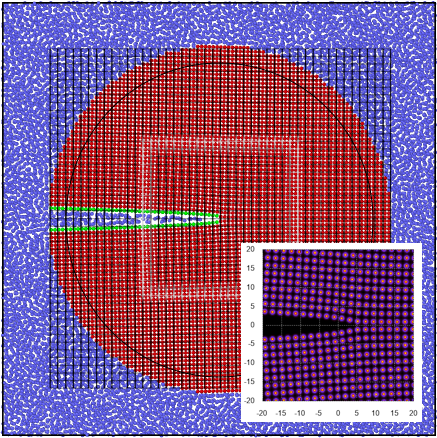}}
\caption{Crack in FCC solid (bulk red, surface green) with infiltrating fluid (blue).
The solid atoms with prescribed displacements are in the annulus outside the black circle.
The mesh for constructing continuum fields is shown in black with contour for evaluating the $\Jb$-integral in gray.
The inset shows the fluid-solid potential energy near the crack tip as probed with fluid test particle. 
}
\label{fig:crack_fluid}
\end{figure}

To model a solid-fluid system with a single slit crack (refer to \fref{fig:crack_fluid}), we employ an atomistic gold-like face-centered cubic (FCC) solid and a water-like simple particle fluid.
The interatomic interactions are given by the commonly-used Lennard-Jones repulsive-attractive potential:
\begin{equation}
\phi_{ab}(r) = 4 \epsilon_{ab} \left( 
\left(\frac{\sigma_{ab}}{r}\right)^{12} -
\left(\frac{\sigma_{ab}}{r}\right)^{6} 
\right) \ ,
\end{equation}
where $r$ is the distance between a pair of atoms/particles of species $a$ and $b$.
The potential starts from 0 at $r\to\infty$, decreases gradually to a well at $r=2^{1/6}\sigma$ and then steeply rises.
It has two parameters:
$\epsilon_{ab}$, which determines energy well depth, and $\sigma_{ab}$, which determines the (soft) radius of the atoms.
For computational efficiency, the interaction potential is truncated at a specified radius, $r_{\text{c},ab} > \sigma_{ab}$.
These parameters are specified in \tref{tab:parameters}.

We fix the solid parameters, $\sigma_\ss$ and $\epsilon_\ss$, to those of a model of (brittle) Au from previous studies \cite{jones2010construction,jones2010atomistic}.
This model has a lattice constant $a$=4.08 \AA, a Youngs modulus $E$=271.95 GPa, a Poisson's ratio $\nu$=0.361437, and a (dry) surface energy  $\gamma_0$=0.159949 eV/\AA$^2$.
The fluid self-interaction energy $\epsilon_\ff$ is fixed to that of the SPC water model \cite{berendsen1981interaction} (other models have similar values \cite{jorgensen1983comparison}).
The size of the fluid particles and their attraction to the solid surfaces are varied to explore steric and hydrophobicity/surface energy effects.
These physical attributes are controlled by $\sigma_\ff\equiv\eta\sigma_{f}$ and $\epsilon_\sf\equiv\xi\epsilon_\ff$, respectively and independently.
In particular, for $\xi < 1$, the fluid is hydrophobic with respect to the solid's surface, \ie it will preferentially adhere to itself; for $\xi > 1$, it is hydrophilic; and, for $\xi > \epsilon_\ss/\epsilon_\ff \approx 107.4768$, it is reactive, in that a solid atom will bind to fluid particles in preference to other solid atoms.
Also, if $\eta  > 0.65$ (estimated from the inset of \fref{fig:crack_fluid}), the fluid particles are unlikely to access the exposed interstitial sites of the solid.
In addition to $\eta$ and $\xi$, we use the standard LJ non-dimensionalization, \eg for the fluid pressure $p_* = p / ( \epsilon_\ff / \sigma_\ff^3 )$ and for the fluid density $\rho_* = \rho \sigma_\ff^3$.
The LJ phase diagram (see Fig. 3 in \cref{lin2003two}, for example) puts constraints on how the parameters can be varied and have the solid and liquid components remain so.
For widest range of pressures where the fluid component remains liquid, we set the system temperature at $T_* = k_B T / \epsilon_\ff = 1.2$ and restrict the density to the range $\rho_* \in [0.55,0.95]$.

\begin{table}
\centering
\begin{tabular}{|l|l|l|}
\hline
\multicolumn{1}{|c|}{$\sigma$ [\AA]} & \multicolumn{1}{|c|}{$\epsilon$ [eV]} & \multicolumn{1}{|c|}{$r_c$ [\AA]}  \\
\hline
$\sigma_\ss =$ 2.59814680      & $\epsilon_\ss =$ 0.72427859 & $r_{c,\ss} =2.1 \sigma_\ss$ \\
$\sigma_\ff = \eta \sigma_\text{f}$   & $\epsilon_\ff =$ 0.00673893 & $r_{c,\ff} =2.5 \sigma_\ff$ \\
$\sigma_\sf = \frac{1}{2} (\sigma_\ss+\sigma_\ff) $ & $\epsilon_\sf = \xi \epsilon_\ff$ & $r_{c,\sf} =2.5 \sigma_\sf$  \\
\hline
\end{tabular}
\caption{Parameters for a gold-like solid (s), with mass $m_\text{s} =$ 196.97 amu, and water-like fluid (f), with mass $m_\text{f} =$ 18.015 amu.
The solid parameters $\sigma_\ss$, $\epsilon_\ss$ and $r_{c,\ss}$ are from \cref{jones2010construction}, the fluid parameter $\epsilon_\ff$ is from \cref{berendsen1981interaction}, the $\sigma_\sf$ value is from Lorentz-Berthelot (LB) mixing, and the remaining $r_c$ values are standard.
The dimensionless variables $\eta$ and $\xi$ control the size of the fluid particles and the hydrophobicity of the fluid with respect to the solid surface.
Nominally $\sigma_\ff = \sigma_\text{f} \equiv 3.16555530$ \AA\ (SPC \cite{berendsen1981interaction}), $\rho_* = 0.8$ \cite{berendsen1981interaction}, and $\xi =\sqrt{\frac{\epsilon_\ss}{\epsilon_\ff}} \approx 10.367$ ($\epsilon_\sf = \sqrt{\epsilon_\ss \epsilon_\ff}$, LB).
}
\label{tab:parameters}
\end{table}

To construct the atomistic crack with infiltrating fluid shown in \fref{fig:crack_fluid}, we created an FCC lattice in a cylindrical region $R < 22 a$ and a fluid at a specific density outside this.
The slit crack was created by deleting bonds across $X_1 < 0$ in the solid.
To obtain the $\Jb$-integral as a function of $K$-loading, $J(K)$, we first displace solid atoms in an outer annulus $R \in [19,22] a$ via  \eref{eq:uLEFM} with a $K$-load increment $0.02 \mu \sqrt{\AA}$.
These atoms are fixed and the remainder are allowed to move.
After an initial energy  minimization, we simulated isothermal (Nos\'e-Hoover) dynamics for 40 ps, at $T_*=6.5$, to accelerate diffusion and allow the fluid to infiltrate crack in a timely fashion, then another 40 ps,  to ramp down to $T_*=1.2$, and lastly 40 ps at $T_*=1.2$.
Finally, we quench the solid component of the system to connect to previous results based on energy minimization \cite{jones2010construction,jones2018atomic} and the $T_*=0$ mechanical properties of the solid.
From these atomic configurations we obtain the necessary continuum fields via a Irving-Kirkwood/Hardy procedure described in full in \cref{zimmerman2010material,jones2010construction,jones2018atomic}.
To insure we sufficiently sample the degenerate fluid and surface states, we report the averages from 200 replica systems.

\section{Results} \label{sec:results}

\begin{figure}[h!]
\centering
{\includegraphics[width=\figwidth]{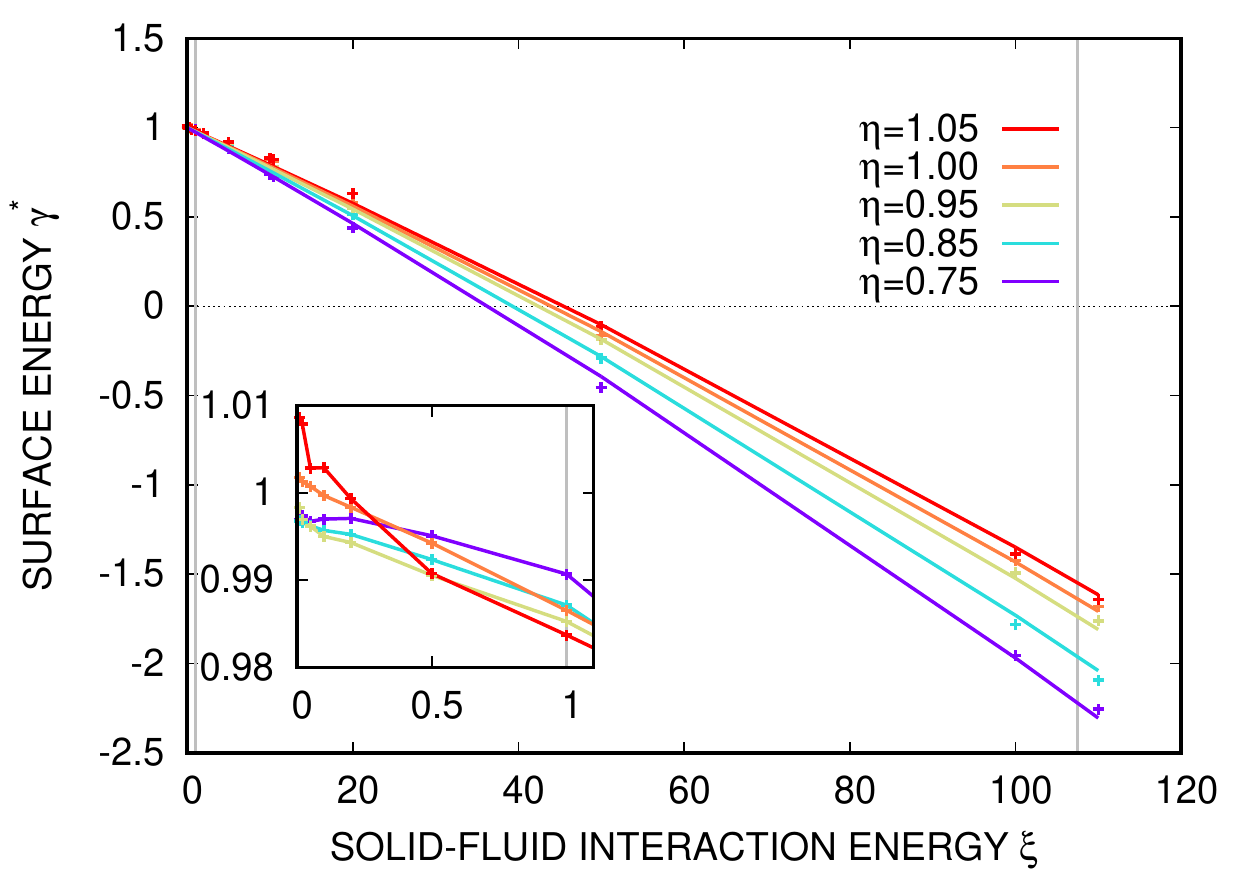}}
\caption{Surface energy as a function of $\xi$, the dimensionless parameter which sets the solid-fluid interaction energy,  and fluid particle size parameter $\eta$.
}
\label{fig:surface_energy}
\end{figure}

A number of preliminary property calculations were necessary.
Since we vary the fluid particle size and want to control the pressure, we estimated the pressure equation-of-state (EOS), $p=p(\rho,T)$, assuming simple fluid behavior at fixed temperature $T$.
We fit isothermal data to a power law to obtain a power-law for pressure $p_* =  p_0 \rho_*^b$ as a function of density $\rho_* \in [ 0.55,  0.95]$, where $p_0 = 9.7012$, $b=4.5295$ and $p_* \in [ 0.646842 , 7.689968]$.
With this EOS, we control pressure for various $\sigma_\ff$ by adjusting the fluid density
\footnote{ Specifically,
$\rho = \frac{\sigma_\ff^3}{\epsilon_\ff} \left(\frac{p_*}{p_0} \right)^\frac{1}{b} $ determined the number of fluid particle in the system.
}.
Also, since the fracture toughness depends on the surface energy $\gamma$, we estimated $\gamma$ via the usual method employing a periodic system with a solid slab and a contacting dense fluid.
With reference to \fref{fig:surface_energy},
for a hydrophobic fluid, $\xi < 1$, the surface energy is effectively constant, and for $\xi > 1$ it fits  reasonably well to: 
$ \gamma^* \equiv \gamma / \gamma_0 =  1.00000 + ( 6.25000  -5.36554 \eta +  1.74062 \eta^2 ) ( -0.05082 \xi  -0.00007 \xi^2 )$.
Note that (a) $\gamma \sim \eta^2$, \ie proportional to packing density of the fluid, and (b) for $\gamma < 0$ a few layers of liquid (the thickness governed by the range of the potential $r_c \sim \eta$) crystallize on the surface of the solid, and (c) we neglect any pressure dependence of $\gamma$.

\begin{figure}[h!]
\centering
{\includegraphics[width=\figwidth]{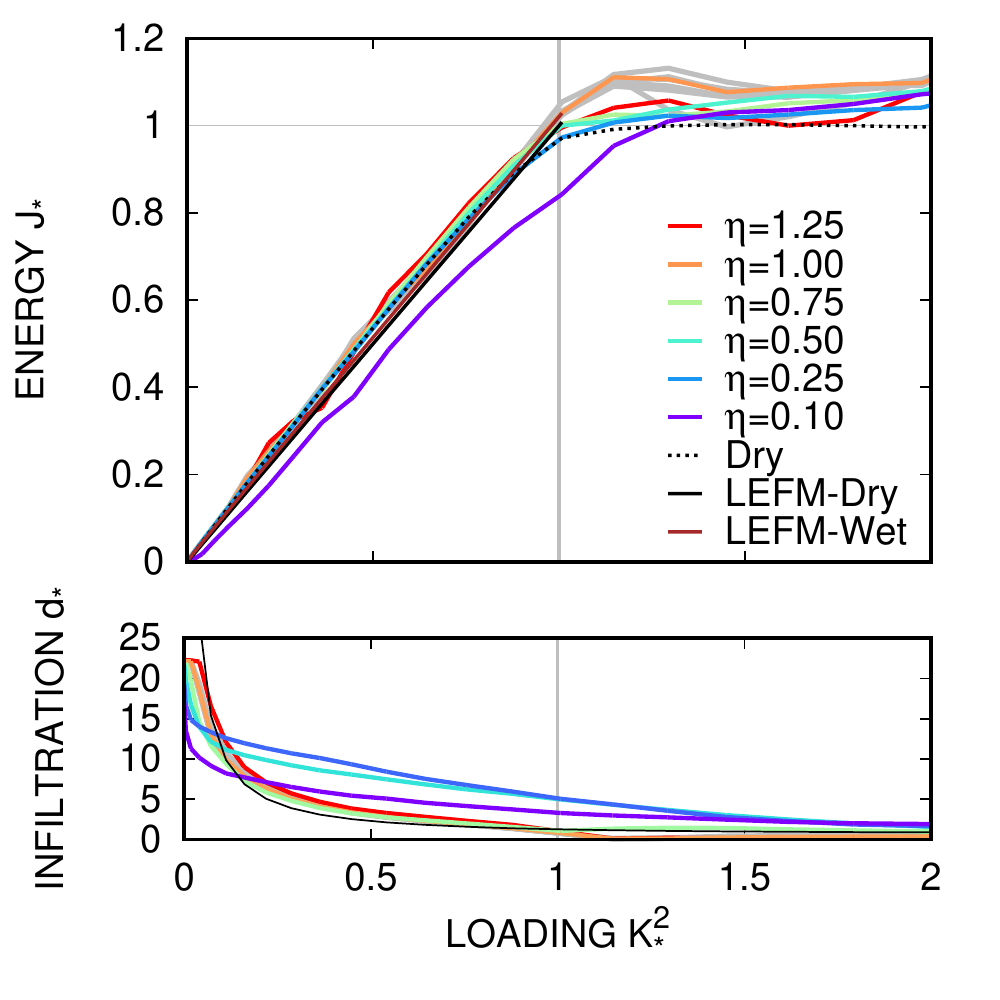}}
\caption{Steric effects on the $\Jb$-integral and infiltration distance $d_*$ controlled by $\eta$ at fixed nominal pressure and solid-fluid interaction energy, $\xi \approx 10.367$. 
The underlying gray lines correspond to the results for the full range of fluid pressures at the nominal $\eta=1$.
In the lower panel, the black line is given by the geometry of the crack and the soft radius of the fluid particles for $\eta = 1$ and the closest approach distance is non-dimensionalized by $a$. 
}
\label{fig:J_sigma}
\end{figure}

We now explore steric and pressure effects at the nominal surface energy.
As expected from theory, in \fref{fig:J_sigma} $J_* \equiv J(K^2)/2 \gamma_0$ is linear up to fracture $K_* \equiv K/\sqrt{2\gamma_0 E^*} \approx 1$ (with a consistent slope across all cases $\eta>0.10$); and pressure has a only modest effect of increasing the slope of the effective $J(K^2)$ curve since the fluid pressures are small compared to the solid's elastic modulus (at the nominal pressure, $p_* = 2.826$, $\Delta K/K_c \approx 0.01$).
For the range of fluid sizes $ 0.25 \le \eta \le 1.00$ the $J$ curves are nearly identical to that calculated for the dry system; however, there are apparent changes in the critical $J_c$ and $K_c$ values at the extreme fluid particle sizes.
For fluid particles greater than the interstitial size, $\eta > 0.50$, the calculated closest approach of the fluid particles to the crack tip $d_*=\min_{\alpha\in\Fc} \| \xb_\alpha \| /a$ compares well to the geometric expectation given by \eref{eq:uLEFM} and the (soft) radius of the fluid particles ($\Fc$ is the group of fluid particles and the crack tip is at the origin)
\footnote{
The continuum solution \eref{eq:uLEFM} gives
$\jump{\ub} = \frac{K (1+\kappa) \sqrt{|X_1|}}{\mu \sqrt{2 \pi}}$, on $X_1 < 0$.
This leads to closest approach of a fluid particle to the crack tip $d = \frac{1}{2} \varrho \left(1 + \frac{r_f^2}{\varrho^2} \right) + r_s$ for $r_f \ge \varrho$ and $d = r_f + r_s$ otherwise, where $\varrho = \frac{(1+\kappa)^2 K^2}{4 \pi \mu^2} $ is the radius of curvature of crack, and $r_f$, $r_s$ are the soft radii of the fluid and solids atoms.
}.
In fact, for these cases, (a) there is a slight increase in $d_*(K^2)$ with $\sigma_\ff$, and (b) near $K^2 = 0$, $d_*$ is constant until the crack opens sufficiently and this transition point depends on $\sigma_\ff$.
For $\eta \le 0.50$, the fluid particles can infiltrate the lattice and do so in preference to diffusing to the tip.
This was observed directly and can be inferred from $d_*(K^2)$.
For $\eta=0.10$ this infiltration is so pervasive (and not localized at the tip) as to soften the lattice and toughen the material relative to the pure solid, as is apparent from the $J(K^2)$ data.
For $\eta = 1.00$, both the critical $J$ and $K$ are significantly higher than  in the dry case.
For $\eta = 1.25$, the critical $J$ and $K$ are higher than in the dry case but the critical $J$ less than in the $\eta = 1.00$ case and yet the critical $K$ is higher.
This trend does not correlate with the variation of $\gamma$ with $\eta$ nor $d_*$ with $\eta$.

\begin{figure}[h!]
\centering
{\includegraphics[width=\figwidth]{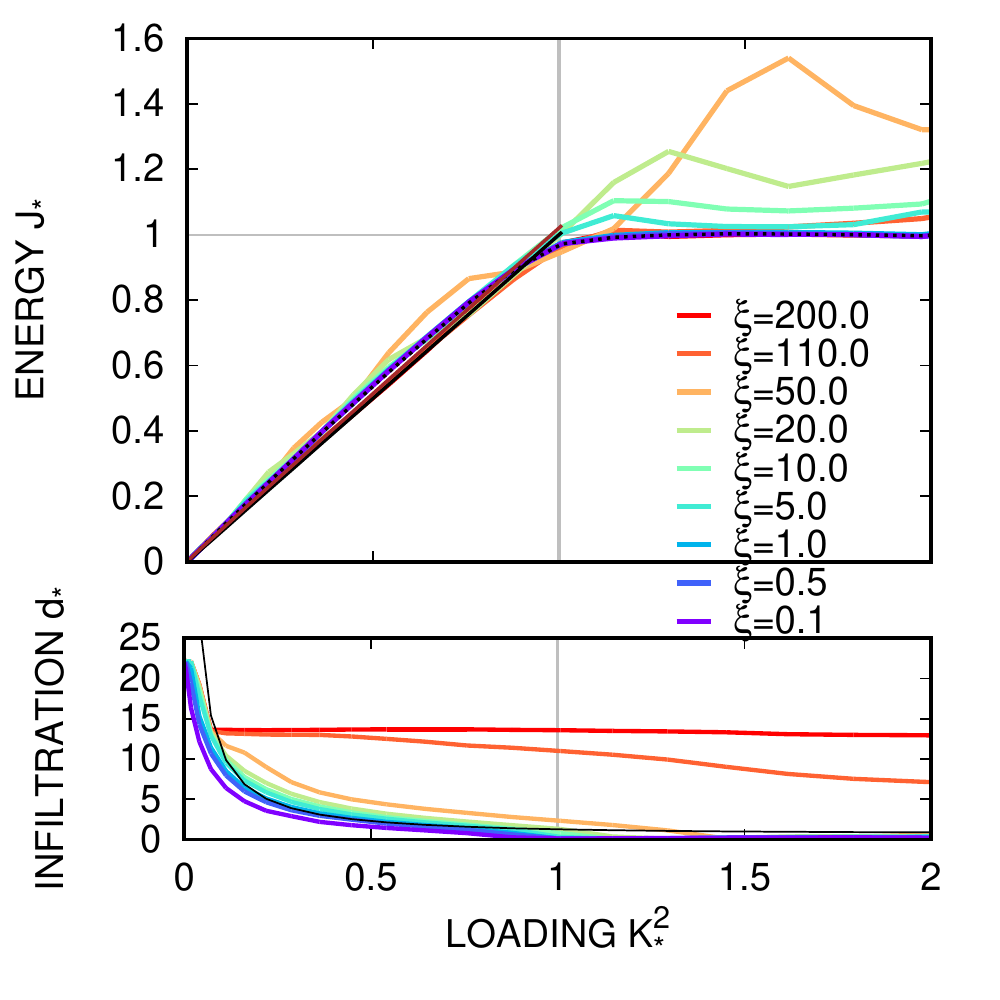}}
\caption{Hydrophobicity effects on the $\Jb$-integral and infiltration distance controlled by $\xi$ at fixed fluid size $\eta = 1.0$.
In the lower panel, the black line is given by the geometry of the crack and the soft radius of the fluid particles and the distance is scaled by $a$.
}
\label{fig:J_epsilon}
\end{figure}

The changes in apparent fracture toughness with changing fluid:solid interactions $\epsilon_\sf$ are less subtle.
\fref{fig:J_epsilon} shows that if the fluid is hydrophobic ($\xi < 1$) the fracture toughness is also unchanged, as expected, since the surface energy is effectively the same.
On the other hand,
as \fref{fig:surface_energy} demonstrates, the surface energy is reduced by hydrophilic fluids ($\xi > 1$) and yet \fref{fig:J_epsilon} clearly shows that both the critical $J$ and $K$ increase with $\xi$.
Our conjecture is that adsorbed fluid monolayers transmit the solid-fluid binding energy between the opposing crack faces at high $\epsilon_\sf \sim \xi$.
This is especially apparent at $\xi=50.00$ where the crack takes on a distinct cusp-like profile and the initial bump in the $J(K^2)$ was observed to be an infiltration, not crack propagation, event.
Generally, higher $\epsilon_\sf$ inhibits the infiltration of fluid into the crack (\ie higher $d_*(K^2)$, with the extreme $\xi > 100$ cases being where unstructured alloys of the solid and liquid species form obstructions that shield the crack tip from contact with the (reactive) fluid and $J(K^2)$ is essentially coincident with the dry case.

\section{Discussion} \label{sec:discussion}

Through atomistic simulation we found that the effects of a hydrophilic fluid can  lead to a trend in fracture toughness that is contrary to Orowan's hypothesis that toughness of brittle solids decreases with the decrease of surface energy due to adsorbed species.
The effect was most pronounced at approximately five times the nominal solid-fluid interaction energy (above which the fluid reacts with the solid).
We conjecture that without a direct mechanism for return of energy of adhesion to crack tip, the transmission of binding forces through atomically thin layers of fluid can dominate and create the observed trend.
We found that for a reasonable range of parameters, steric effects were relatively minor, except for fluid particles small enough to infiltrate and soften the solid lattice.
Of course, this may change under extreme fluid pressures.
Lastly, we observed that highly reactive fluids can obstruct fluid access to the crack tip.

\section*{Acknowledgments}
We are grateful for technical discussions with Jonathan A. Zimmerman.
This work was supported by the LDRD program at Sandia National Laboratories, and its support is gratefully acknowledged.
Sandia National Laboratories is a multimission laboratory managed and operated by National Technology and Engineering Solutions of Sandia, LLC., a wholly owned subsidiary of Honeywell International, Inc., for the U.S. Department of Energy's National Nuclear Security Administration under contract DE-NA-0003525.
The views expressed in the article do not necessarily represent the views of the U.S. Department of Energy or the United States Government.


\end{document}